\documentstyle[floats,aps,prl,epsfig,twocolumn,amstex]{revtex}

\begin{document}
\bibliographystyle{/usr/share/texmf/tex/latex/revtex/prsty}
\draft
\wideabs{
\title{Further comment on pion electroproduction and the axial form factor}

\author{
V.~Bernard,${^1}$ N. Kaiser,${^2}$ Ulf-G.~Mei{\ss}ner${^3}$
}

\address{$^{1}$Laboratoire de Physique Th\'eorique,
Universit\'e Louis Pasteur, F-67084 Strasbourg Cedex, France}
\address{$^{2}$Physik Department T39, TU M\"unchen, D-85747 Garching, Germany} 
\address{$^{3}$Institut f\"ur Kernphysik (Theorie), Forschungszentrum J\"ulich
D-52425 J\"ulich, Germany}

\maketitle

\begin{abstract}
We show that a recent claim \cite{haber} that one cannot extract the nucleon weak axial 
form factor $G_A (t)$ from charged pion threshold electroproduction is incorrect.
Thus previous calculations remain valid and threshold charged pion electroproduction
experiments can  indeed be used to determine $G_A (t)$, and they
should certainly be pursued.
\end{abstract}
\pacs{PACS number(s): 13.60.Le, 11.40.Ha, 14.20.Dh}
}
\narrowtext
\noindent
In a recent paper, Haberzettl~\cite{haber} claims that one cannot extract
the weak axial form factor $G_A (t)$ of the nucleon from threshold electroproduction
of charged pions as first stressed by Nambu and collaborators~\cite{NLS}.
He argues that previous derivations of the relationship between $G_A (t)$
and the electromagnetic structure of the Kroll-Ruderman contact terms
are based on incomplete evaluations of the relevant PCAC expressions and
that, if all mechanisms are taken into account, the dependence of the
pion electroproduction amplitude on $G_A (t)$ vanishes.

\noindent
Let us follow the derivation of~\cite{haber} and show were it goes 
wrong. (We will later discuss
the method underlying that paper.) The matrix element of the weak axial current 
between nucleon states is indeed given by:
\begin{equation}
j_A^\mu =-\bar u_f (p') \gamma_5 \Biggl( \gamma^\mu G_A (t) + 
\frac {(p-p')^\mu}{2m} G_p (t)
\Biggr) \frac{\tau}{2} u_i (p)~, \label{eq:ja}
\end{equation}
with $t= (p'-p)^2$ the invariant momentum transfer squared, $m$
denotes the nucleon mass, $G_A (t)$ and $G_P (t)$ are the axial and the
induced pseudoscalar form factor, respectively. 
Exploiting the chiral Ward identity of QCD relating the divergence of the
axial current to the pseudoscalar density one gets to leading order
for $G_P$
\begin{equation}
G_p (t)  = \frac {4m g_{\pi NN } F_\pi}{M_\pi^2 -t} + {\cal O}(t^0)~,
\label{eq:gp}
\end{equation} 
which is the well known leading pion pole contribution to $G_p$ expressed in
terms of the pion--nucleon coupling constant $g_{\pi NN}$ and the weak
pion decay constant $F_\pi$.
The corrections to this result have also been obtained 
but are of no interest for the following discussion.
Consequently, the nucleon matrix element of the axial current,
Eq.(\ref{eq:ja}), thus  contains a pion pole dominated part.
This is a direct consequence of the spontaneous chiral symmetry
breaking of QCD as first observed by Nambu.
This axial current matrix element is unambiguous and unique, and so is
the axial current. The key observation is now that there is no need, as has been
done in~\cite{haber},  to split the axial current into two pieces and
introduce 
a so--called ``conserved weak part'' $\hat j_{A,W}$ at the prize of introducing
some unphysical and problematic singularity at $t = 0$.
This splitting leads to the wrong claim in~\cite{haber}. 
This observation was already made by Guichon~\cite{pierre}.
The axial current can indeed be represented as in fig.1 of~\cite{haber}, i.e. by a
pion--pole term and remainder, but then
$\hat j_{A,W}$ is an unconserved quantity which contains all but the
pion-pole contribution and which together with  $\hat j_{A,H}$
{\em does not contain any unphysical
singularity}. This is particularly important since now $\hat j_{A,H}$
describes the creation of a pion of mass $M_\pi$ out of the vacuum, with 
the coupling operator $-F_\pi (p'-p)^\mu$ and associated normalized ``form factor''
1 and  {\it not}
 $M_\pi^2/t$, and the subsequent propagation of the pion and its
final absorption in the nucleon, where we have used the same phrasing
as in~\cite{haber}. This is nothing but the well known QCD relation:
$\langle 0 | A_\mu |\pi \rangle \sim F_\pi  p_\mu$  which is easily recovered in chiral
perturbation theory.
At that point it is fairly easy to see that in the derivation of eq.(19)
of ref.\cite{haber}, which is the main point of Haberzettl's note, the divergence
of the last term in eq.(14) of that paper which is proportional to 
$\hat j^\mu_\pi$ will {\it not} contribute
to the photoproduction amplitude $\cal {M}$ as  claimed by Haberzettl.
Indeed it does not lead to a term of the form
\begin{equation}
\sim \frac {M_\pi^2}{q^2 -M_\pi^2} M_{\rm int}^\nu~,
\end{equation}
but rather to a structure of the type 
\begin{equation}
\sim \frac {q^2}{q^2 -M_\pi^2} M_{\rm int}^\nu~.
\end{equation} 
where $ M_{\rm int}^\nu$ is the interaction current defined in eq.(12)
of~\cite{haber} (see also fig.2 of~\cite{haber}). 
The difference in these last two expressions can be traced back
to the ``normalized form factor'' of $\hat j_\pi^\mu$. It will thus contribute to
the last term $\bar u_f {\cal W}^\nu u_i \epsilon_\nu$ which vanishes
in the soft pion limit. Thus $q_\mu J^{\mu \nu}_{A, \gamma} \epsilon_\nu$
will {\it not} involve this contribution from $\hat j^\mu_\pi$ contrary to 
what Haberzettl claims. It will {\it indeed
vanish even when} $q \to 0$.\footnote{The third term 
and fifth term in eq.(14) of~\cite{haber} which also depend on $\hat j^\mu_\pi$
will only lead to a modified expression for ${\cal W^\nu}$.}  
Thus there is one term less in the photoproduction
amplitude, the one corresponding to the
last diagram in fig.2 of~\cite{haber}.
We have:
\begin{equation}
q_\mu J_{A,\gamma}^{\mu,\nu} \epsilon_\nu = \frac {f_\pi M_\pi^2}{q^2 -
M_\pi^2} ({\cal M}-{\cal M}_{\rm int}) + \bar u_f {\cal W^\nu} u_i \epsilon_\nu
\end{equation}
where ${\cal M}_{\rm int}$ contains the Kroll-Ruderman contact term among others. 
But as shown by Haberzettl, this contact term is just given by $Q_\pi j_A^\nu
\epsilon_\nu$ in the soft pion limit, so that one naturally gets back to 
eq.(19) of~\cite{haber} 
as it should. Note that ${\cal W^\nu}$ does not verify eq.(20) anymore, 
the term $Q_\pi j^\nu_{A,W}$ 
is replaced by 
\begin{equation}
\frac {q^\nu}{t-M_\pi^2} \,  \gamma_5 \,
g_{\pi NN} \, \tau
\end{equation}
and one has additional terms coming from other
diagrams contributing to ${\cal M}_{\rm int}$. 
The important point is that now the axial form factor $G_A (t)$ enters only 
via $Q_\pi j_A^\nu \epsilon_\nu$ in the soft pion limit as in all previous 
calculations and contrary to Haberzettl's claim. There is 
no more cancellation of the axial form factor which was
 solely coming from this ad hoc splitting of the 
axial current into parts containing unphysical singularities at $t=0$. 

We also remark that the method of \cite{haber} applied to this particular
problem  is extremely clumsy. The correct relation
based on the chiral Ward identities of QCD can be obtained 
much more easily by making use of
an effective Lagrangian as it is known since decades.\footnote{To quote a famous
Harvard physicist, the method of \cite{haber} appears to us as  an
exercise in self-torture.} Indeed, we have used such
methods to derive the one--loop corrections to the NLS~\cite{NLS} low-energy theorem,
which is claimed to be erroneous by Haberzettl, in \cite{BKMax}. 
The steps in \cite{BKMax}
are extremely simple to follow and they also show that Haberzettl's metaphysical
remarks about the relation of his results to the ones obtain in chiral perturbation 
theory are unfounded. Current algebra  is nothing but the first term in a systematic
expansion about the chiral limit of QCD and thus must lead to the same result
as a corresponding lowest order chiral perturbation theory calculation. Of course,
one has to be aware of possible pitfalls - some current commutators are not
well defined and making extra assumptions can lead to incorrect results, the
best example being the incorrect low--energy theorem for neutral pion photoproduction
off nucleons~\cite{BGKM}. In fact, the NLS low--energy theorem was exactly missing
the additional ``axial radius correction'' found in \cite{BKMax} because the
smoothness assumption of going from massless to massive pions does not commute
with taking the derivative of the electric dipole amplitude $E_{0+}^{-} (k^2)$
with respect
to the photon virtuality $k^2$ for $k^2$ tending to zero. Needless to say that
using the method of \cite{haber} it seems impossible to us to recover such an
intricate correction. Independent of this correction, the effective Lagrangian
method directly leads to the NLS result and the author of \cite{haber} has to
proof that that derivation is also incorrect. Parenthetically, we are utterly
amazed that the referees of \cite{haber}  did not even perform this extremely
simple check. 

We have thus shown that the claim of Haberzettl that
one cannot extract the nucleon weak axial 
form factor $G_A (t)$ from threshold pion electroproduction in fact is
wrong. It is vital to perform further precise charged pion electroproduction
experiments to not only get a better determination of the nucelon axial
form factors but also have an alternative method to measure the pion charge
radius, see e.g. \cite{BKMpi}.


\end{document}